\documentstyle[sprocl,psfig]{article}

\def\Journal#1#2#3#4{{#1} {\bf #2}, #3 (#4)}

\def\PLB{{\em Phys. Lett.}  B}
\def\PRL{\em Phys. Rev. Lett.}

\def\be{\begin{equation}}
\def\ee{\end{equation}}
\def\bea{\begin{eqnarray}}
\def\eea{\end{eqnarray}}

\begin{document}

\title{NEUTRINO-INDUCED NEUTRON SPALLATION AND 
THE SITE OF THE $r$-PROCESS}

\author{YONG-ZHONG QIAN}

\address{Physics Department, 161-33, 
California Institute of Technology, \\Pasadena, CA 91125
\\E-mail: yzqian@citnp.caltech.edu} 


\maketitle\abstracts{All of the actinides and 
roughly half the natural abundance of 
elements with mass number $A>70$ 
come from the rapid neutron capture process, or the $r$-process.
If the $r$-process, as suggested by many, occurs deep in
a supernova, then it is under the influence of an intense
neutrino flux. Here we discuss the effects of both charged-current
and neutral-current neutrino interactions on the $r$-process.
We show that the multiple-neutron emission induced by both kinds 
of neutrino interactions can affect the observed $r$-process 
abundance pattern significantly. In particular, we find that
five nuclei below the $r$-process abundance peak at $A\sim 195$
may be entirely attributed to the neutrino-induced neutron
spallation processes following the $r$-process freeze-out.
Furthermore, the deduced neutrino fluence agrees with the
conditions in the recent supernova $r$-process model. These
results strongly argue that the $r$-process occurs in supernovae.
They also provide a sensitive probe for the conditions at the 
supernova $r$-process site.}

\section{Introduction}
It is well known that almost all of the $^4$He nuclei in our
universe were created in the Big Bang. We also know that stars
shine by burning H into successively heavier elements. However,
nuclear physics tells us that $^{56}$Fe is the most tightly bound
nucleus. No more nuclear binding energy can be released to power
the star by burning Fe. Therefore, elements heavier than $^{56}$Fe
have to be made in some other processes. One such process is the
rapid neutron capture process, or the $r$-process for short.
This process is responsible for all of the actinides and 
about half the natural abundance
of elements with mass number $A>70$. Here
we are mainly interested in the $r$-process nuclei near $A\sim 195$.
Typical $r$-process elements in this mass region are Os, Pt, and Au.

The $r$-process mechanism is rather simple because there are only
three types of reactions involved. A nucleus can increase its mass
number by capturing neutrons. If the temperature is high enough,
photodisintegration competes with neutron capture. In any case,
if a nucleus becomes too neutron rich, it can also $\beta$-decay.
By definition, neutron capture is much faster than $\beta$-decay
during the $r$-process. Furthermore,
the $r$-process mechanism 
can explain the observational data
quite well. There are two prominent peaks at $A\sim 130$ and 195,
respectively, in the observed solar $r$-process abundance pattern.
The existence of these two peaks can be explained as follows.
\vfill
\eject
At the beginning of the $r$-process, there are some seed nuclei
and lots of neutrons. The seed nucleus then captures 
neutrons and moves towards the neutron-drip line. However, at
some point, the binding energy of the next neutron to be captured
becomes so small that it will be quickly disintegrated by the
photons available in the $r$-process environment. At this 
so-called ``waiting-point,'' the nucleus must $\beta$-decay to a 
new nuclear species before further neutron capture can proceed.
Through such a series of neutron capture and $\beta$-decay, a
distribution of progenitor nuclei far away from $\beta$-stability
is produced on the $r$-process path. Clearly, the progenitor
abundance at a given charge number $Z$ is piled up at the
corresponding waiting-point nucleus before it $\beta$-decays,
and the more slowly it $\beta$-decays, the more abundant it will
be. Due to the extreme stability of the closed neutron shells,
the $\beta$-decay rates for the progenitor nuclei at the magic
neutron numbers $N=82$ and 126 are extremely small. Consequently,
abundance peaks are formed at the progenitor nuclei with these 
magic neutron numbers. After neutron capture stops, these
progenitor nuclei successively $\beta$-decay to stability and
give rise to the peaks at $A\sim 130$ and 195 in the observed
solar $r$-process abundance pattern.

Although the above $r$-process theory succeeds in explaining
the main features of the observed solar $r$-process abundance
pattern, it does not elaborate on where the $r$-process occurs.
Perhaps the most popular site for the $r$-process is the 
interior of a supernova. However, a crucial step to identify supernovae
as the $r$-process site is to find something
hidden in the observed $r$-process abundance pattern that has the 
signature of a supernova. Probably 
one of the best known signatures 
of a supernova is its powerful neutrino emission,
and as we show in the following discussion,
we can indeed extract some fingerprints of neutrinos
from the observed $r$-process abundance pattern. In particular,
we find that five nuclei below the abundance peak at $A\sim 195$
may be entirely produced by the neutrino-induced neutron spallation
processes after the rapid neutron capture stops (i.e., following
the $r$-process freeze-out). Our discussion is organized as follows.
In Sec. 2, we briefly describe the emission of 
supernova neutrinos and the characteristics of their interactions
with the neutron-rich progenitor nuclei produced during the 
$r$-process. In Sec. 3, we discuss the effects of neutrino-induced
neutron spallation on the $r$-process and identify the five nuclei
that are most sensitive to these effects. We discuss the 
implications of our results for the $r$-process site in Sec. 4.

\section{Neutrino Emission and Interactions in Supernovae}
A supernova occurs when the core of a massive star
collapses into a compact neutron star with a mass of 
$\sim 1\ M_\odot$ and a radius of $\sim 10$ km. The gravitational
binding energy of the final neutron star is $\sim 10^{53}$ erg.
Due to the high temperatures and densities encountered during the 
collapse, the most efficient way to release this energy is to emit
neutrinos and antineutrinos mainly through electron-positron pair
annihilation. In fact, because of the intense elastic 
neutral-current scatterings on free nucleons for all 
neutrino species, even neutrinos have to diffuse out of the
neutron star on a timescale of $\sim 10$ s, as confirmed by the
detection of neutrinos from SN1987a.~\cite{87a} Since these
neutrinos are in thermal equilibrium with the neutron star matter 
and with each other for most part of the diffusion process, the
individual neutrino luminosities are about the same and have an
average value of $L_{\nu}\sim 10^{51}$ erg s$^{-1}$.

However, the average neutrino energies are very different. This
is because these neutrinos have different abilities to exchange
energy with matter. The elastic neutral-current scatterings on
free nucleons, which govern the diffusion of all neutrinos,
essentially cost neutrinos no energy because the nucleon rest
mass is much higher than the typical neutrino energies. All
neutrino species can exchange energy with matter through 
scatterings on electrons. However, only $\nu_e$ and $\bar\nu_e$
are energetic enough to have charged-current capture reactions
on free neutrons and protons, respectively. Therefore, 
$\nu_{\mu(\tau)}$ and $\bar\nu_{\mu(\tau)}$ have the weakest
ability to exchange energy with matter and decouple from thermal
equilibrium at the highest temperatures and densities as they
diffuse towards the neutron star surface. Correspondingly, they
have the highest average energy. Between $\nu_e$ and $\bar\nu_e$,
the $\nu_e$ have more chances to exchange energy with matter
because there are more neutrons than protons inside a neutron
star. Consequently, the $\nu_e$ decouple at the lowest 
temperatures and densities, and have the lowest average energy.
The average $\bar\nu_e$ energy lies in the middle. Typically,
the average neutrino energies are 
$\langle E_{\nu_e}\rangle\approx 11$ MeV,
$\langle E_{\bar\nu_e}\rangle\approx 16$ MeV, and
$\langle E_{\nu_{\mu(\tau)}}\rangle=
\langle E_{\bar\nu_{\mu(\tau)}}\rangle\approx 25$ MeV.

With such powerful neutrino emission, neutrino interactions with
the progenitor nuclei can be very important if the $r$-process
indeed occurs in supernovae. The charged-current $\nu_e$ capture
reaction and the inelastic neutral-current scatterings for 
$\nu_{\mu(\tau)}$ and $\bar\nu_{\mu(\tau)}$ are of particular interest.
For the average supernova neutrino energies given previously,
charged-current capture reactions are energetically forbidden
for $\nu_{\mu(\tau)}$ and $\bar\nu_{\mu(\tau)}$. However, these
neutrinos dominate the inelastic neutral-current scatterings
because they have the highest average energy and the inelastic
neutral-current scattering cross section of interest is strongly 
energy dependent. On the other hand, the charged-current $\bar\nu_e$
capture reaction is unimportant because it changes a proton into
a neutron, and all the allowed transitions are Pauli blocked for
the extremely neutron-rich progenitor nuclei made in the $r$-process.
Obviously, the neutrino interaction rates scale with the neutrino flux
(i.e., $\propto L_\nu/r^2$). For the progenitor nuclei at $A\sim 195$,
the charged-current $\nu_e$ capture rate is 
$\sim 8$ s$^{-1}$ and the total 
neutral-current scattering rate is $\sim 12$ 
s$^{-1}$ at a distance of $r=100$ km from the neutron star
for a luminosity of
$L_\nu=10^{51}$ erg s$^{-1}$ per neutrino species.

With an average energy of $\sim 10$ MeV, the
$\nu_e$ mainly put the daughter nucleus to the excited states at 
$\sim 10$ MeV above the ground state of the parent nucleus
through the charged-current capture reaction. A typical
ground state energy difference between the parent and daughter nuclei
is also $\sim 10$ MeV. So typically the daughter nucleus has an
excitation energy of $\sim 20$ MeV above its ground state. 
Because the progenitor nuclei made in the $r$-process are extremely
neutron rich, their daughter nuclei
deexcite almost exclusively through neutron
emission. The neutron binding energy ranges from $\sim 3$ MeV for the
progenitor nuclei to $\sim 8$ MeV for the stable nuclei. On the 
average, $\sim 4$ neutrons are emitted after each $\nu_e$ capture 
reaction, and the branching ratios for emitting different numbers of 
neutrons can be estimated using statistical techniques.~\cite{qian}
As for the $\nu_{\mu(\tau)}$ and $\bar\nu_{\mu(\tau)}$, they
can excite the progenitor nucleus to the states near the neutron
emission threshold or to those at $\sim 15$ MeV above the ground state
through the inelastic neutral-current scattering.
On the average, $\sim 2$ neutrons are emitted after each 
inelastic neutral-current scattering, and various neutron
emission branching ratios can be estimated as in the charged-current
case.

\section{Effects of Neutrino-Induced Neutron Spallation on the 
$r$-Process}
During the dynamic phase of the $r$-process, neutrino-induced
neutron spallation is unimportant because the 
typical rates for neutron capture and photodisintegration are
faster than those for the neutrino reactions by orders of 
magnitude. When the neutron number density drops below a
critical level, rapid neutron capture stops and the progenitor
abundance pattern freezes out. Ideally, this progenitor abundance
pattern would lead to the observed abundance pattern. However,
standard $r$-process calculations without neutrino effects
have a major deficiency. For example, these calculations greatly
underproduce the nuclei at $A\sim 182$--187 below the abundance
peak at $A\sim 195$. This is because these calculations give
a generic progenitor abundance pattern that lacks the 
corresponding progenitor nuclei.~\cite{friedel} 

Due to the competition between neutron capture and 
photodisintegration, it turns out that the progenitor nuclei
on the $r$-process path have roughly the same neutron binding
energy.~\cite{friedel} In general, at a given charge number $Z$
the neutron binding energy decreases with increasing neutron
number $N$, and at a given $N$ it increases with increasing $Z$.
Therefore, the same neutron binding energy is reached at a
larger $N$ as $Z$ increases. Due to the existence of the closed
neutron shell at $N=126$, the neutron binding energy for a
given $Z$ decreases very slowly as $N$ increases towards 126.
As a result, when $Z$ increases from 63 to 64, $N$ jumps from
118 to 124 for the progenitor nuclei on the $r$-process path,
and the progenitor nuclei at $A\sim 182$--187 are not produced
in significant abundance during the $r$-process. This then leads
to the deficiency in the corresponding mass region of the final
abundance pattern.

However, if neutrino reactions are significant after the 
$r$-process freeze-out, the final abundance at a given $A$ will 
receive contributions from progenitor nuclei at higher mass
numbers due to 
the neutrino-induced neutron spallation. We refer to this effect
of neutrino interactions on the $r$-process abundances
as ``neutrino postprocessing.''  With neutrino postprocessing,
the final abundance $Y_f(A)$ at mass number $A$ is given by
\begin{equation}
Y_f(A)=\sum_{n=0}^{n_{\rm max}}P_nY_{\rm pro}(A+n),
\end{equation}
where $Y_{\rm pro}(A+n)$ is the progenitor abundance at mass number
$A+n$, and $P_n$ is the probability for emitting a total of $n$ 
neutrons during postprocessing. Provided that we have the right form
of these postprocessing neutron emission probabilities, it is
possible that we can reproduce the observed abundance pattern
everywhere despite the deficiency in the progenitor abundance
pattern. 

The postprocessing neutron emission probability $P_n$ can be 
calculated as follows. First of all, we note that
there are different ways
to emit a total of $n$ neutrons. For example, a total of two
neutrons can be emitted by one event emitting two neutrons or
by two events each of which emits one neutron. The occurrence
of an event emitting $i$ neutrons is governed by the Poisson
distribution
\begin{equation}
P(N_i)={\langle N_i\rangle^{N_i}\over N_i!}
\exp(-\langle N_i\rangle),
\end{equation}
where $\langle N_i\rangle$ is the average number of events 
emitting $i$ neutrons. The postprocessing neutron emission 
probability $P_n$ can then be obtained by listing the ways
to emit a total of $n$ neutrons and summing up the corresponding
probabilities calculated from the Poisson distributions.~\cite{qian}

In Eq. (2), $\langle N_i\rangle$
depends on the average number of
neutrino interactions during postprocessing and the branching ratio
for emitting $i$ neutrons after each neutrino interaction. The
neutrino interaction rates and the associated
neutron emission branching ratios have been
discussed in Sec. 2. More details were given elsewhere.~\cite{qian}
The average number of neutrino interactions also depends on the
neutrino fluence 
\begin{equation}
{\cal{F}}=\int_{t_{\rm fo}}^\infty{L_\nu(t)\over r(t)^2}dt,
\end{equation}
where $t_{\rm fo}$ is the time at the $r$-process freeze-out.

The postprocessing neutron emission probabilities for the progenitor
nuclei at $A\sim 195$ are plotted for 
${\cal{F}}/(10^{47}\ {\rm erg\ km}^{-2})=0.015$, 0.030, and
0.060 in Fig. 1 (hereafter, the values of $\cal{F}$ are always
given in unit of $10^{47}$ erg km$^{-2}$).
Two features of these probabilities are worth
noticing. The first feature is that for 
all three chosen neutrino fluences,
the probability for emitting zero neutron is the largest. This
is because for these fluences, the average number of neutrino
interactions is $\sim 1$ or less. Consequently, according to
Poisson statistics, the probability for no interaction
at all is large. The second feature is that these probabilities
have bumps at $n=4$ and 5. This is because the neutron emission
characteristics of charged-current and neutral-current neutrino
interactions are very different. On the average, $\sim 4$ 
neutrons are emitted after each charged-current interaction, 
twice as many as in the neutral-current case. The superposition
of these two kinds of interactions then leads to the bumps
at $n=4$ and 5.

\begin{figure}
\centerline{\psfig{file=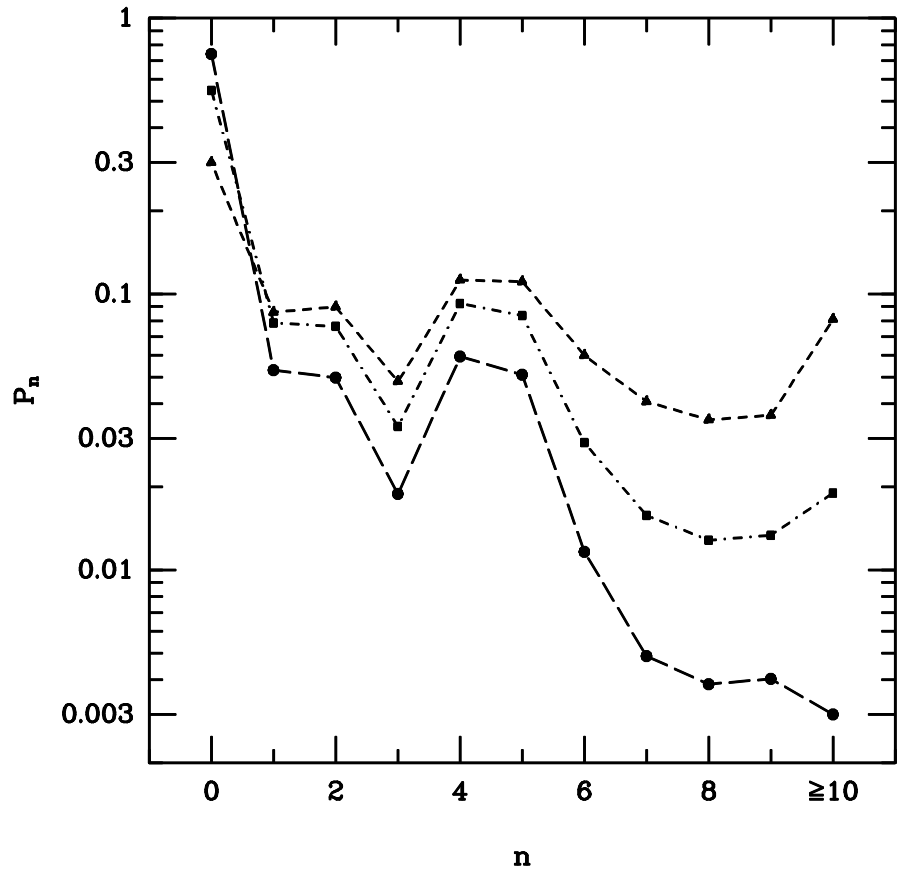,width=4.7in}}
\caption{Postprocessing neutron emission probabilities
for the progenitor nuclei at $A\sim 195$. The points connected
by the long-dashed, dot-dashed, and short-dashed lines are for
neutrino fluences of
${\cal{F}}=0.015$, 0.030, and 0.060, respectively.}
\end{figure}

These postprocessing neutron emission probabilities, which only depend
on the neutrino fluence, are needed to 
test if the observed abundances at $A\sim 182$--187 can be produced
by neutrino postprocessing.
The progenitor abundances at $A>187$ are also needed for this test.
However, since the observed abundances at $A>187$ should be
reproduced with the same neutrino fluence, a given number of these
progenitor abundances can be obtained from the same number of
constraints on the final abundances after postprocessing [see
Eq. (1)]. Therefore, we are trying to find one neutrino fluence
that can fit a number of abundances at $A\leq 187$ in this test.

Our results are shown in Fig. 2. In this figure, 
the data points, some with
error bars, give the observed solar $r$-process abundances. 
The solid line gives the progenitor abundance pattern, and the
dashed line gives the final abundance pattern after neutrino 
postprocessing. These results correspond to a neutrino fluence
of ${\cal{F}}=0.015$. The results for $A=183$--187 are highlighted
in the inset. For the best-fit fluence of ${\cal{F}}=0.015$,
the observed abundances at these five mass numbers are reproduced
within $1\ \sigma$, although the progenitor abundance pattern
is deficient at these mass numbers. Because the final abundances
at these mass numbers entirely come from neutrino postprocessing,
they are the fingerprints of neutrinos on the $r$-process
abundance pattern.

\begin{figure}
\centerline{\psfig{file=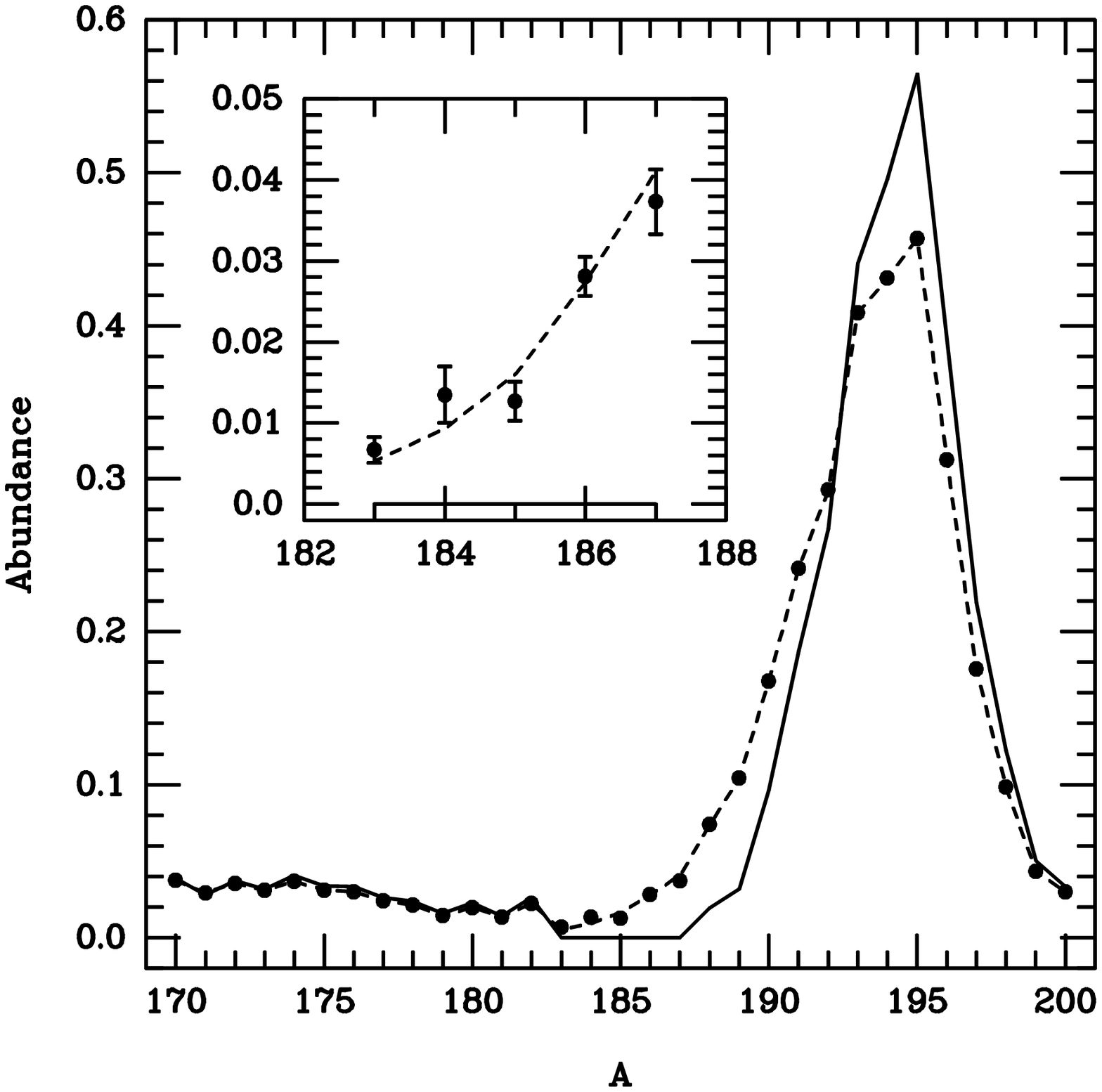,width=4.7in}}
\caption{Neutrino postprocessing results in the $A\sim 195$
region for a fluence of ${\cal{F}}=0.015$. 
The data points, some with error bars, give the observed
solar $r$-process abundances. The solid (dashed) line gives
the abundance pattern before (after) neutrino postprocessing.}
\end{figure}

\section{Implications for the Site of the $r$-Process}

It was suggested that the deficiency of the standard
$r$-process calculations in the mass region of $A=183$--187
could result from the deficiency of nuclear mass models for the
extremely neutron-rich nuclei far away from stability.~\cite{friedel}
It was shown that a similar deficiency below the abundance peak
at $A\sim 130$ no longer exists after the strength of the $N=82$
closed neutron shell is quenched.~\cite{chen} However,
quenching the strength of closed neutron shells may not be a general
way to eliminate this kind of deficiencies. As explained
in Sec. 1, the abundance peaks at $A\sim 130$ and 195 result
from the sharp contrast in the $\beta$-decay rates for the 
nuclei with and without closed neutron shells. Clearly, quenching
the strength of closed neutron shells tends to diminish this contrast.
Therefore, it may not be always possible
to make the deficiencies below the abundance peaks disappear
by quenching the strength of closed neutron shells without
changing the distinct features of the abundance peaks at
the same time.

We have shown in Sec. 3 that for a neutrino fluence of
${\cal{F}}=0.015$, the deficiency of the standard $r$-process
calculations in the mass region of $A=183$--187 can be remedied
completely by neutrino postprocessing without 
quenching the strength of closed neutron shells.
Furthermore, we can show that this fluence is 
consistent with the conditions in the recent supernova
$r$-process model. The postprocessing neutrino fluence
is approximately determined by the neutrino flux at the 
$r$-process freeze-out and the typical timescale over which
the neutrino flux significantly decreases. The neutrino
flux decreases either due to the decay of the neutrino
luminosity or due to the increasing distance of the $r$-process
material from the neutron star. A typical decay timescale
for the neutrino luminosity is $\sim 3$ s, whereas the
$r$-process material is ejected with a dynamic timescale
of $\sim 0.1$--1 s.~\cite{stan} In addition, the $r$-process
typically freezes out at $\sim 300$--1000 km in the recent
supernova $r$-process model of Woosley {\it et al}.~\cite{stan}
For a luminosity of $L_\nu=10^{51}$ erg s$^{-1}$ per neutrino
species, these conditions correspond to a neutrino fluence
of ${\cal{F}}\sim 0.01$, which is very close to the best-fit
value found in Sec. 3.

To summarize, we find that
the deficiency
of the standard $r$-process calculations in the mass region
of $A=183$--187 can be remedied completely
by neutrino-induced neutron spallation following the $r$-process
freeze-out. In addition, the neutrino fluence needed for this
postprocessing agrees with the conditions
in the recent supernova $r$-process model.
These results strongly argue that the $r$-process
occurs in supernovae. Furthermore, they provide a sensitive
probe for the conditions at the supernova $r$-process site
because the freeze-out radius and the dynamic timescale
for the $r$-process are now severely constrained by the 
postprocessing neutrino fluence.

\section*{Acknowledgments}
This work was done in collaboration with W. C. Haxton, 
K. Langanke, and

\noindent P. Vogel. Y.-Z. Qian was supported by the 
David W. Morrisroe Fellowship
at Caltech.

\end{document}